\documentclass{egpubl}

\ifpdf \usepackage[pdftex]{graphicx} \pdfcompresslevel=9
\else \usepackage[dvips]{graphicx} \fi
\PrintedOrElectronic

\usepackage{amsmath}
\usepackage{graphicx}
\usepackage{amssymb}

\title[MarchingSurfaces] {Marching Surfaces: Isosurface Approximation using G$^1$ Multi-Sided Surfaces}

\author[Gustavo Ch{\'a}vez, Alyn Rockwood] {Gustavo Ch{\'a}vez, Alyn Rockwood}

\begin{document}

\maketitle

\begin{abstract}
Marching surfaces is a method for isosurface extraction and approximation based on a $G^1$ multi-sided patch interpolation scheme. Given a 3D grid of scalar values, an underlying curve network is formed using second order information and cubic Hermite splines. Circular arc fitting defines the tangent vectors for the Hermite curves at specified isovalues. Once the boundary curve network is formed, a loop of curves is determined for each grid cell and then interpolated with multi-sided surface patches, which are $G^1$ continuous at the joins. The data economy of the method and its continuity preserving properties provide an effective compression scheme, ideal for indirect volume rendering on mobile devices, or collaborating on the Internet, while enhancing visual fidelity. The use of multi-sided patches enables a more natural way to approximate the isosurfaces than using a fixed number of sides or polygons as is proposed in the literature. This assertion is supported with comparisons to the traditional Marching Cubes algorithm and other $G^1$ methods.
	
\end{abstract}

\section{Introduction}

Isosurface extraction, approximation and display is fundamental in visualization for volume data, e.g., in scientific visualization. Marching cubes \cite{lorensen1987marching} has been the \emph{de facto} standard for indirect volume rendering since its inception in 1987; major scientific software still uses it despite disadvantages such as serrated edges, angularities, linear precision, and particularly large polygon count. 

Our goal is to create an isosurface representation and rendering method that is visually appealing, yet accurate, useful for mobile computing and collaboration, and ultimately able to leverage recent parallel hardware capabilities in a parallel version that handles large datasets in accordance with \cite{johnson2004top}. This paper introduces the method and demonstrate its correctness and compressibility. We selected high-order surfaces that provide smoother reconstruction and exhibit fewer rendering artifacts than existing methods.

The first step is to reduce the 3D gridded dataset to a a representation on the three independent axial planes. That allows us, to reformulate the problem of finding isosurfaces to that of finding isolines. Next, we deduce the connectivity of the isolines. We simplify this process by allowing only cell-by-cell cube-to-plane intersections bounded by only four different types of high-order surfaces, i.e. with three, four, five and six sides. By using neighboring cell information we provide a more accurate reconstruction that avoids some of the well-known ambiguity cases of marching cubes \cite{Nielson91} and allows us to have curve network with $C^1$ continuity. Geometric continuity is a major advantage of using high-order composite surface reconstruction. We provide a set of conditions that enable multi-sided surface patches to join with $G^1$ continuity at the cross boundaries.

\textbf{Related work}. This paper builds on a significant amount of previous research in indirect volume rendering and surface approximation methods using high-order interpolation surfaces. These methods can be classified according to the type of surfaces in which they represent the isosurface or implicit function of interest. 

The seminal work of \cite{gallagher1989efficient} on three-dimensional finite element simulations and other coarse volumes uses bi-cubic polynomials in the form of Ferguson patches. Their approach yields surfaces and surface normals that approximate high-order threshold surfaces directly at the corner values of the cells of interest. One drawback is the noticeable discrepancies between the surface shape and light source shading near the bi-cubic boundaries. The authors call this issue "result smoothing", which comes from the fact that Ferguson patch is not the ideal type of surface for this purpose.

The work of \cite{sorokina2007local} describes a scheme based on cubic $C^1$ splines on type-6 tetrahedral partitions of volumetric grids. The authors solve an optimization problem to find the appropriate coefficient combinations that generate a boundary description of the tetrahedron; operators are able to provide an error bound and construct the tri-variate function that will serve as the basis for volume reconstruction. This technique falls into the class of quasi-interpolation methods for tri-variate splines. Related methods are \cite{kalbe2008hardware}, \cite{Kalbe2009}, and \cite{marinc2012interactive}; they leverage the relatively low total degree of the splines to construct a hardware accelerated implementation for real time reconstruction of isosurfaces. Tri-linear interpolation is a common choice of high-order approximation methods; unfortunately, smoothness of adjacent splines is not a trivial task. Furthermore, these schemes require high-order derivatives that have to be approximated. 

Trimmed surfaces of triangular rational cubic Bezier patches is the basis of \cite{Theisel:2002:CGF}, the author introduces it as a direct extension of marching cubes providing a topologically exact approximation of contours. It leverages $G^{1}$ continuity at the expense of a global re-parametrization step at the end of its method. Although trimmed representations allows the use of a single type of patch, it adds unnecessary complexity compared to our method that naturally chooses the best type of patch cell-by-cell.

Other related methods that provide $G^1$ continuity are \cite{farin2012agnostic}, \cite{liu2008approximate}, \cite{nowacki97} and \cite{Mann92surfaceapproximation}; these methods achieve continuity of the adjoining control points by an adjustment to the boundary curve that ensure continuous tangent plane everywhere. While we make use of the same tangent plane requirement, we introduce the concept of tangential ribbon which puts no restriction to the underlying boundary curve network. This is the as well last step of our method, but with the difference that the computation is local with respect to every cell.

The need of a lightweight volume rendering for restricted environments such as the web browser or mobile devices is becoming evident in recent publications such as \cite{Rodriguez12}, \cite{Jacinto12}, and \cite{Chavez2013}. The standard choice for indirect volume rendering is still marching cubes; in this paper we advocate for the high resolution representation here presented as the method of choice due to its compact representation and the economy of storing and transmitting complex surfaces that are broken down into multi-sided composite surfaces with $G^1$ continuity.
\section{Boundary Curve Network}

\begin{figure}[ht!]
\centering
\includegraphics[width=.7\linewidth]{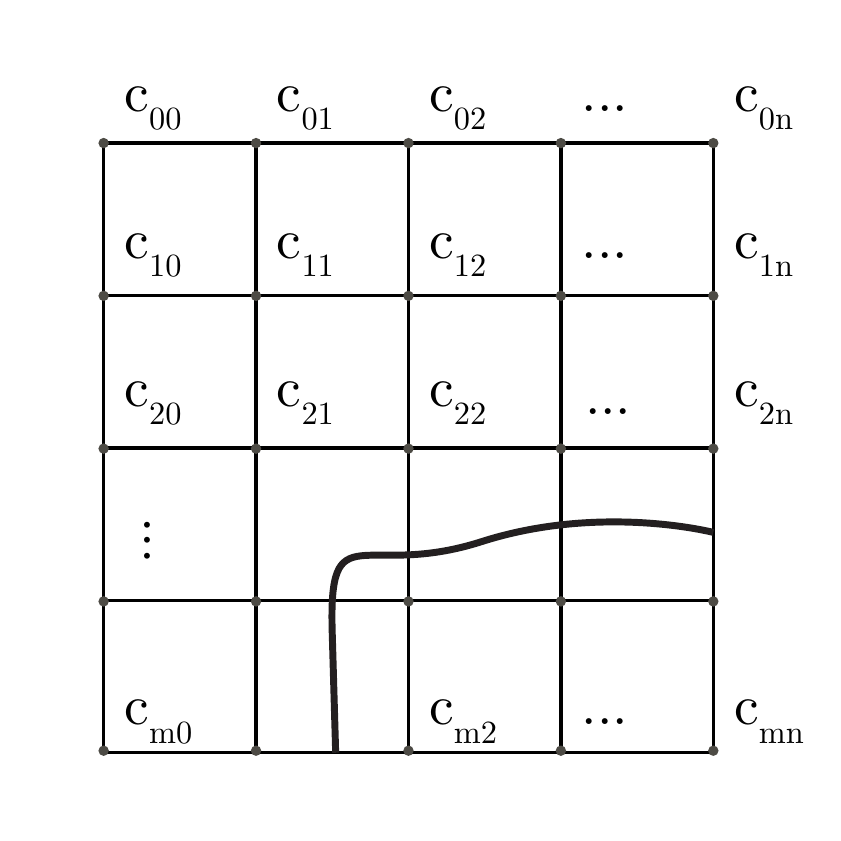}
\caption{Axial plane with function values at each grid point $F_{ij} =c $, and an isoline representing a boundary curve.}
\label{fig:images/2D_grid.pdf}
\end{figure}

\begin{figure}[ht!]
\centering
\includegraphics[width=.7\linewidth]{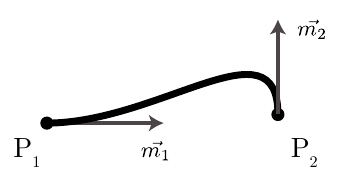}
\caption{A cubic Hermite curve is defined by two points and two tangent vectors. Appropriate choose of tangents between Hermie splines assures C$^1$ continuity.}
\label{fig:images/hermiteCurve.pdf}
\end{figure}

The first step towards marching surfaces is to construct an underlying boundary curve network per each axial plane; namely, \emph{xy, yz} and \emph{xz}. We extract a set of curves that represent \emph{isolines} at a function values \emph{c}, called \emph{isovalue}. This set of points in the domain map to $\left\{ x\in \mathbb{R}:f(x)=c \right\}$ \cite{courant89}, as illustrated in figure~\ref{fig:images/2D_grid.pdf}. We find an approximation of $x:f(x) = 0$ based on cubic Hermite splines (\ref{eq:hermite}); appropriate choose of tangent vectors $m_1$ and $m_2$ (\ref{eq:hermite}) assures global $C^1$ continuity across a connected sequence of curves and a high order parametric interpolation \cite{farin2002}, see figure \ref{fig:images/hermiteCurve.pdf}.

\begin{align}
p(t) = &~(2t^3-3t^2+1)p_0 + (t^3-2t^2+t)m_0 \label{eq:hermite}\\ 
&+ (-2t^3+3t^2)p_1 +(t^3-t^2)m_1 \nonumber
\end{align}

\begin{figure}[!ht]
\centering
\includegraphics[width=.7\linewidth]{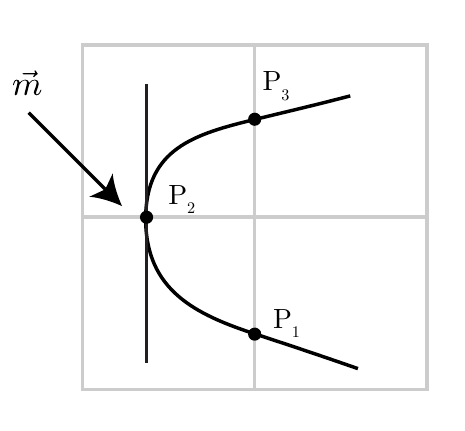}
\caption{Choosing three consecutive points to make use of second order information. The tangent vector $\vec{m}$ at the middle point is computed by fitting a quadratic curve.}
\label{fig:images/fit3points.pdf}
\end{figure}

Each piece of the isoline is a third-degree polynomial specified by two control points and the first derivatives at the start and end points, over the unit interval $(0,1)$; given $p_0$ at $t=0$ and $p_1$ at $t=1$ with $t\in[0,1]$. Control points are placed at lines with zero crossing based on a linear proportion of $f(x)_{ij}$ values at grid points. Tangents are computed by fitting a quadratic curve trough three points at the time over neighboring cells, as shown in figure~\ref{fig:images/fit3points.pdf}.

Parabolas are the first curve of choice \cite{Theisel:2002:CGF} to derive tangents, but considering that isolines should stay as close as possible, we suggest that circular arcs provide maximum entropy (min max of curvature) as derived in \cite{strang2007computational}. It is shown that the best approximation after a straight line is a circular arc; which is the shortest curve with minimal area (\ref{eq:proofMinimalenergy}).

\begin{align} 
& \underset{P(u)}{\text{minimize}} & & \int_0^1 \sqrt{1+(u')^2}\,dx \label{eq:proofMinimalenergy}\\
& \text{subject to}                     & & u(0)=a,\; u(1)=b,\; \int_0^1 u(x)\,dx = A \nonumber
\end{align}

\begin{figure}[!ht]
\centering
\includegraphics[width=.7\linewidth]{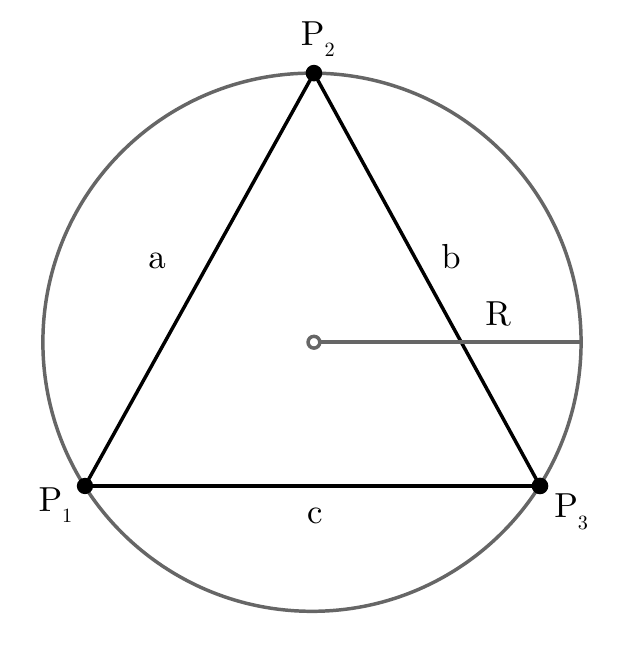}
\caption{Connecting 3 consecutive points to form a triangle in order to find the radius $R$ of the enclosing circle. a is the length between $P_1$ and $P_2$, b between $P_2$ and $P_1$, and c between $P_1$ and $P_3$.}
\label{fig:images/Circlediagram.pdf}
\end{figure}

A circular arc can be fitted by computing the circumradius $R~(\ref{eq:R})$ of the enclosing circle that is form by connecting the three consecutive points of interest with lengths $a$, $b$, and $c$ that form a triangle; then the calculation of its semi-perimeter $s~(\ref{eq:s})$ and area $K~(\ref{eq:K})$ according to Heron's Formula and \cite{johnson1929modern} as illustrated in figure ~\ref{fig:images/Circlediagram.pdf}.

\begin{align}
&s = \frac{1}{2}(a+b+c) \label{eq:s}\\
&K =  \sqrt{(s (s-a) (s-b) (s-c))} \label{eq:K}\\
&R =  \frac{abc}{4K} \label{eq:R}
\end{align}

\begin{figure}[!ht]
\centering
\includegraphics[width=.7\linewidth]{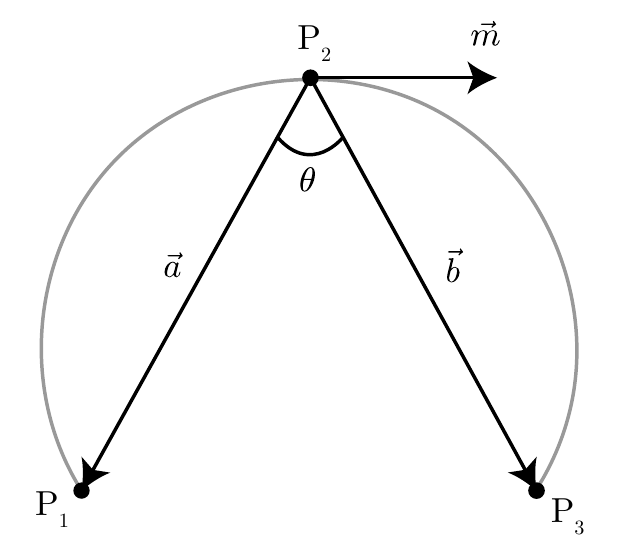}
\caption{Circular arc approximation given 3 points. Vectors $\vec{a~} and \vec{~b}$ are used to find the angle $\theta$ between them, to ultimately define the tangent vector $\vec{m}$ at the middle point $P_2$.}
\label{fig:images/arc.pdf}
\end{figure}

Using the parametric equation of the circle it is possible to find the tangent vector $\vec{m}$ of the middle point, with angle $\theta \in [0,\pi]$ and vectors ${\vec{a} = P_1 - P_2}$, and ${\vec{b} = P_3-P_2}$, see figure~\ref{fig:images/arc.pdf}.

\begin{align}
&\theta = \arccos (\frac{ A \cdot B}{\|A\| \|B\|}) \label{eq:theta} \\
&m = [\frac{-R\sin(\theta)}{R\cos(\theta)}] \label{eq:m}
\end{align}

The construction of a $C^1$ continuous set of Hermite curves using three consecutive points requires precise definitions of tangents at the first and last point. Every point in the isoline is composed by the tuple ${(t_k,P_k)}$ for ${k=1,\ldots,n}$. For $k=1$ tangent is defined by (\ref{eq:firstTangent}) and for $k=n$ by (\ref{eq:lastTangent}); tangents of the middle points, i.e. $k=2,\ldots,n-1$ are defined at (\ref{eq:middleTangent}). An special case has to be defined when the isoline does not form a loop, here we provide tangents definitions for $k=1$ and for $k=n$ (\ref{eq:lastTangentNoLoop}).

\begin{align}
&m_0(P_{n},P_{1},P_{2}) \label{eq:firstTangent} \\
&m_k(P_{k-1},P_{k},P_{k+1}) \label{eq:middleTangent} \\
&m_n(P_{n-1},P_{n},P_{1}) \label{eq:lastTangent} \\
&m_0*(P_{1},P_{2},P_{3}) \label{eq:firstTangentNoLoop} \\
&m_n*(P_{n-2},P_{n-1},P_{n}) \label{eq:lastTangentNoLoop}
\end{align}

The length of the tangents can be adjusted for different fits. When the adjacent points are nearly co-linear we suggest that $1/3$ of the distance between endpoints is an effective factor, which comes from the definition of a cubic Bezier spline, that mimics the legs of its control polygon. As the tangent legs become right angles a factor of $\frac{1}{3}\sin(\theta)$ lengthens them and the curve straightens up, this gives optimal visual fidelity in most of the cases.

\begin{figure}
\centering
\includegraphics[width=.5\linewidth]{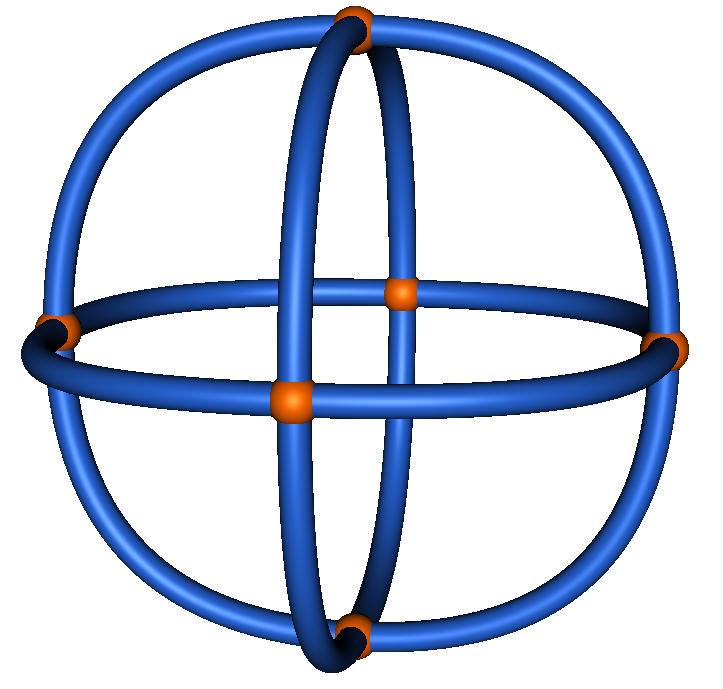}
\caption{Boundary curve network of implicit function $ x^2+y^2+z^2 -1 = 0 $ over [-3, -3, -3] x [3, 3 ,3]. Isovalue = 0.0. In blue are the boundary curves, in orange are the control points. Showing one isoline per axial plane.}
\label{fig:test6caseA}
\end{figure}

\begin{figure}
\centering
\includegraphics[width=\linewidth]{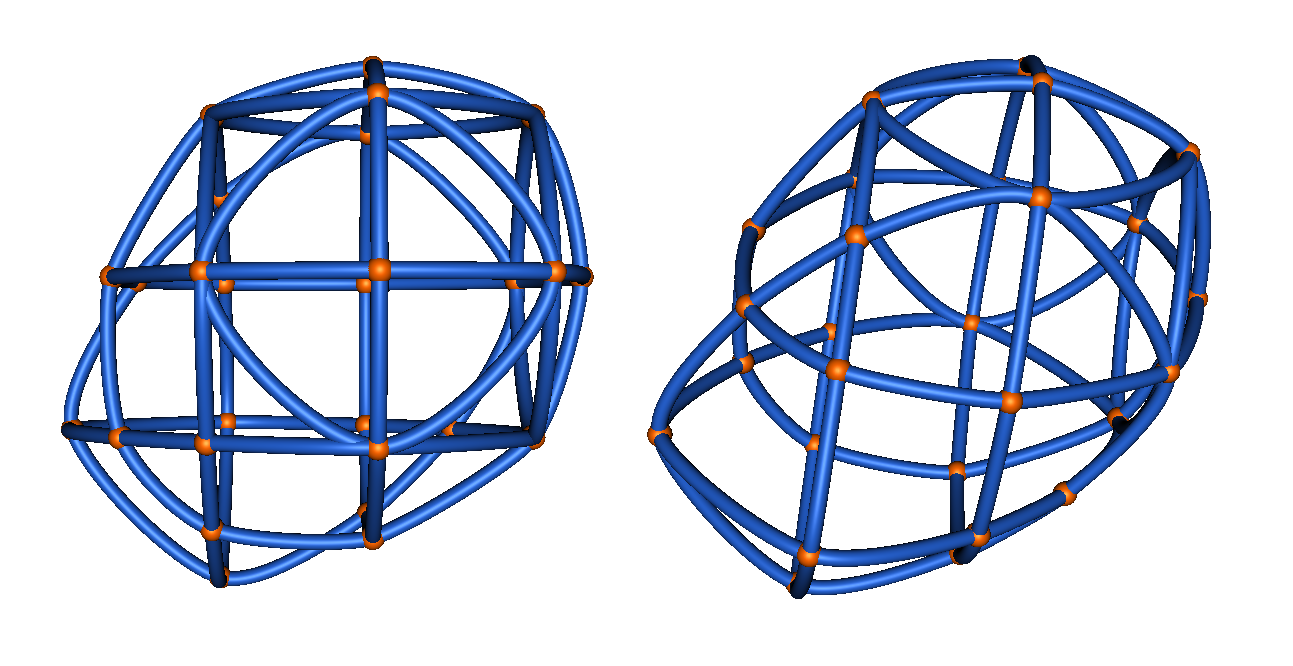}
\caption{Boundary curve network of implicit function constructed of the blend of a sphere and a hyperboloid $ \frac{3}{4}((x+1)^2+(y+1)^2+(z+1)^2-1)+\frac{1}{4}(xyz-3) = 0 $ over [-3, -3, -3] x [3, 3 ,3]. Isovalue = 0.0. Show several isolines per axial plane. Isolines are globally G$^1$ continuous based on Cubic Hermite Splines.}
\label{fig:test6caseB}
\end{figure}

Once points and tangents are defined at every zero crossings in all the cells of the plane, curves can be rendered in each axial plane. Up to this point we have an intermediate visualization method readily available formed out of boundary curves. Figure (\ref{fig:test6caseA}) and (\ref{fig:test6caseB}) illustrate the reconstruction of two implicit functions with its corresponding isovalues on $\mathbb{R}^3$.

\section{Multi-sided surface construction}

\begin{figure}[!ht]\centering
\includegraphics[width=\linewidth]{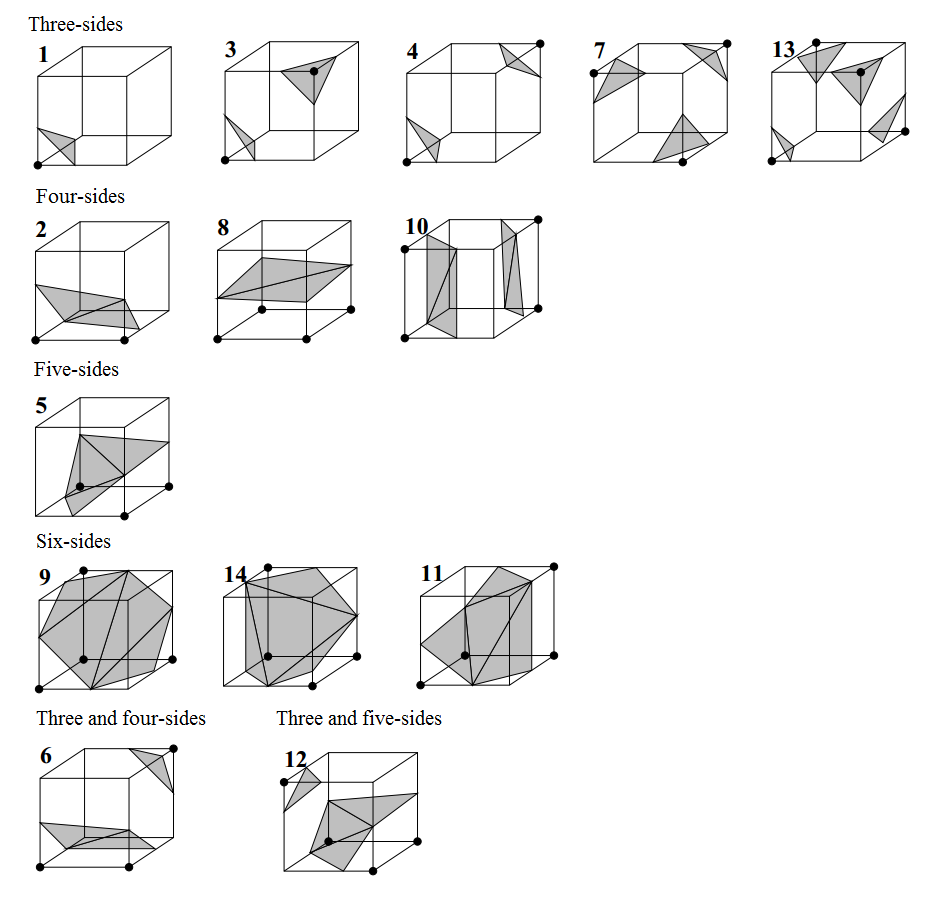}
\caption{Out of the 15 original lookup table cases of the Marching Cubes algorithm, we group the choice of higher order surfaces to four types of three, four, five and six sides.}
\label{fig:images/reductionCases.png}
\end{figure}

The resulting intersections across boundary curves naturally suggest the number of sides of the necessary surfaces to fill in cell-by-cell. As mentioned before, our method selected a type of surfaces flexible enough to handle any number of surfaces sides. If we consider the original look up table of 15 cases \cite{Chernyaev95} of cube-to-plane intersection, we can group all the possible cases by sets of three, four, five and six-sided surfaces only, as summarized in figure \ref{fig:images/reductionCases.png}.

Surface construction is based on \cite{gao2005multi} and \cite{VaradyRS11}. They define an interpolation scheme based on the weighted, discrete least squares solution $\boldsymbol{x} = (x,y,z)$ that minimizes: $ \sum_{i} [(x-x_i)^2+(y-y_i)^2+(z-z_i)^2]W_i(x,y,z)$. The least squares solution slews toward the points with larger weights ${W}_i(u)$. Weight selection for multi-sided patches is based on a parametric map of control polygons on $\mathbb{R}^2$ with the same number of sides \emph{k} as the required surface, called \emph{footprints} and denoted by $U_i$ (figure \ref{fig:images/footprints.pdf}). Boundary curves are then used to map each footprint\rq{s} point $u_i$ into the resulting surface in $\mathbb{R}^3$, denoted by the attribute function $f_i(u_i)$. Weights are given as reciprocal distances from the boundary curves $f_i$ to the footprints $U_i$. The following equations (\ref{eq:fundDef}) give the complete surface construction definition:

\begin{figure}[!ht]
\centering
\includegraphics[width=.7\linewidth]{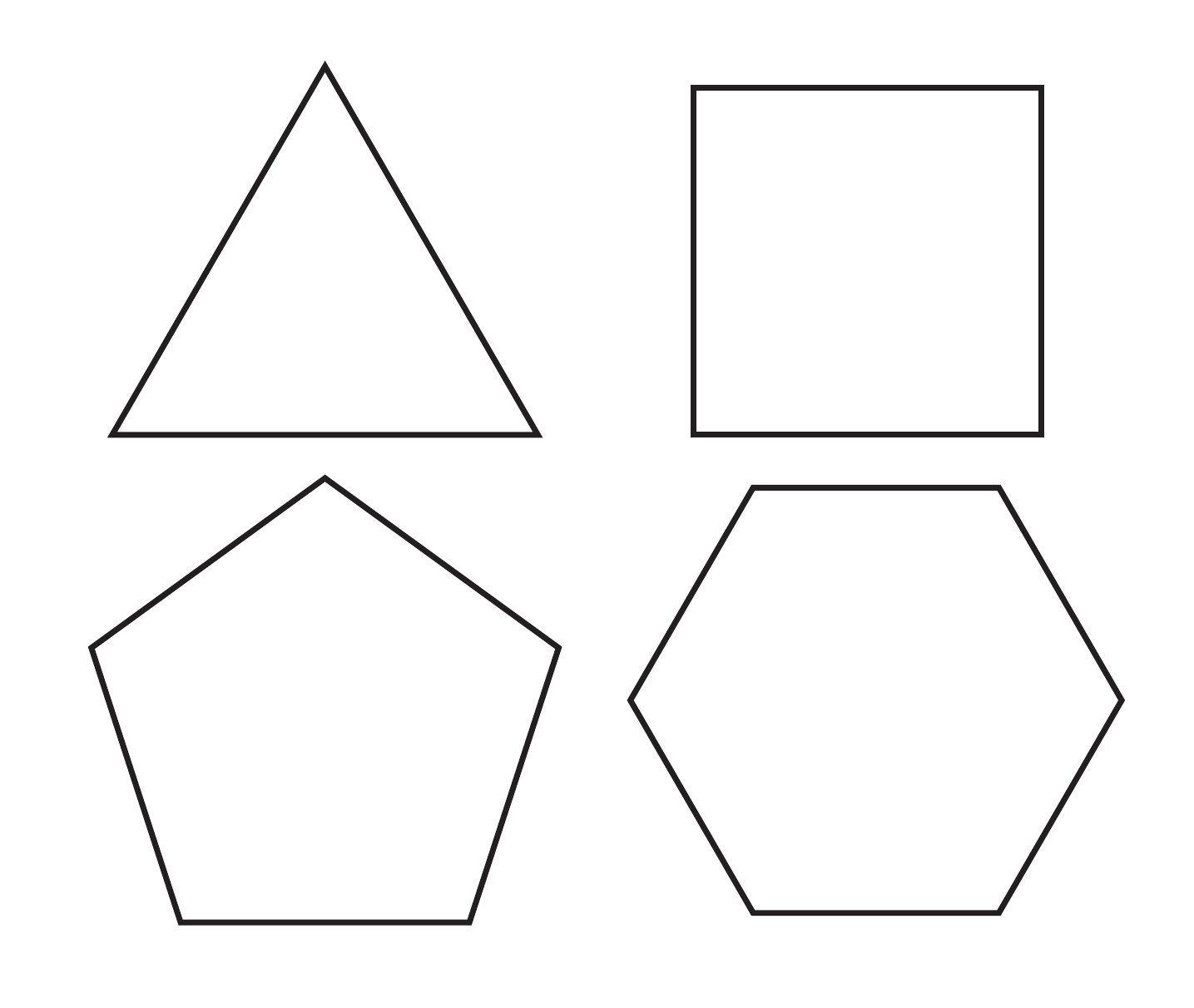}
\caption{Footprints for three, four, five and six-sided surfaces.}
\label{fig:images/footprints.pdf}
\end{figure}

\begin{align}
F(u)=(x,y,z)=\sum_{i}f_i(u_i)\hat{W}_i(u) \label{eq:fundDef}\\
\hat{W}_i(u)={W}_i(u)/\sum_{k}{W}_k(u) \nonumber
\end{align}

The previous surface definition together with a number of unique points per cell allows us to start filling in the regions that will constitute the partial isosurface in a composite and independent manner, from here the selection of the word \emph{marching} to our method, in comparison with the industry standard marching cubes.

\begin{figure}[!ht]
\centering
\includegraphics[width=.8\linewidth]{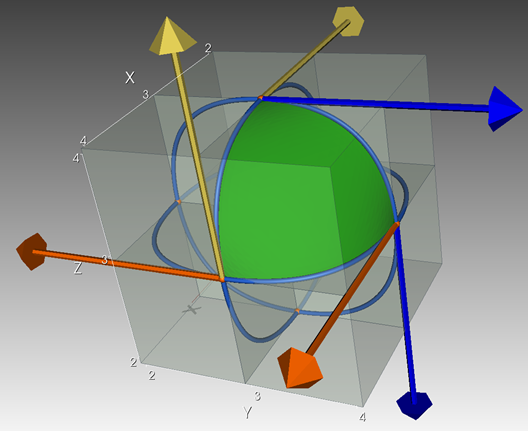}
\caption{Partial reconstruction showing one three-sided surface. Arrows represent the tangent vectors at the adjoin points.}
\label{fig:g0reconA}
\end{figure}

\begin{figure}[!ht]
\centering
\includegraphics[width=\linewidth]{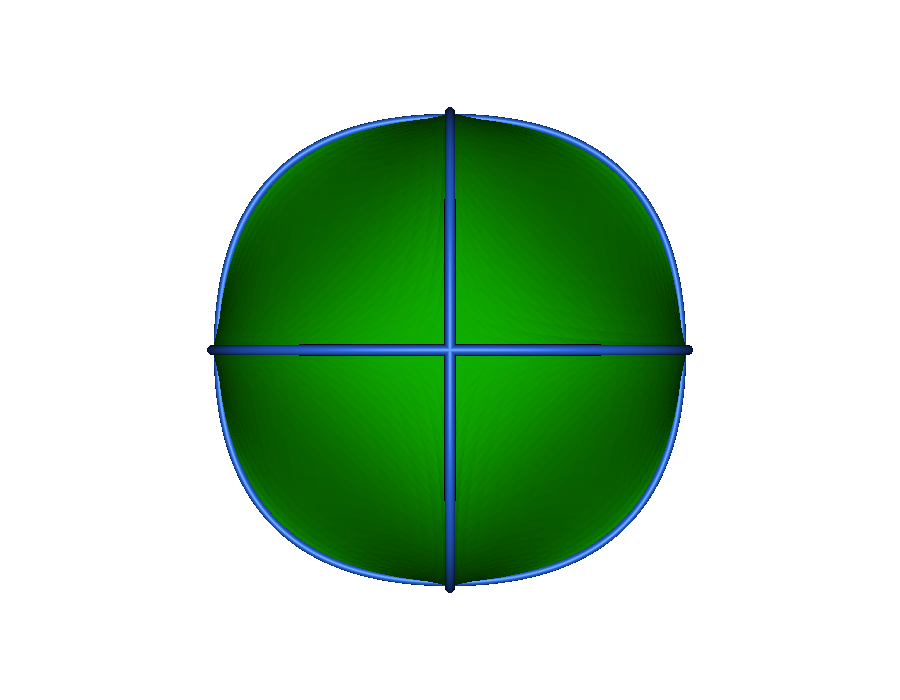}
\caption{Final composite surface of a unit sphere with eight three-sided surfaces and G$^0$ continuity; image also depicts its corresponding boundary curve network.}
\label{fig:g0reconB}
\end{figure}

Depending of the vertex valence, there might be multiple tangents directions to be picked up per adjoining point. Selecting the correct tangent direction is key to get a correct reconstruction. In practice, the implementation could use a dataset that stores the tangent directions and its corresponding axis, or take the dot product between the normalized tangent and the normalized vector between start and end point, selecting the tangent with the smallest angle as the appropriate. Tangent directions already suggest an arrangement of points in loop form, also for practice application, we suggest an additional order selecting a consistent wind direction for all the surface faces. The resulting arrangement will be the in the form of a loop of curves as shown in the three-sided patch at figure \ref{fig:g0reconA}. The complete reconstruction is shown in figure \ref{fig:g0reconB}. The set of neighboring surfaces yield to a fair surface reconstruction with G$^0$ continuity.

\section{G$^1$ continuous composite surfaces}

G$^1$ continuity requires continuous tangent planes between neighboring surfaces, and they are accomplished by \emph{tangential ribbons}. 

\begin{figure}[!ht]\centering
\includegraphics[width=.8\linewidth]{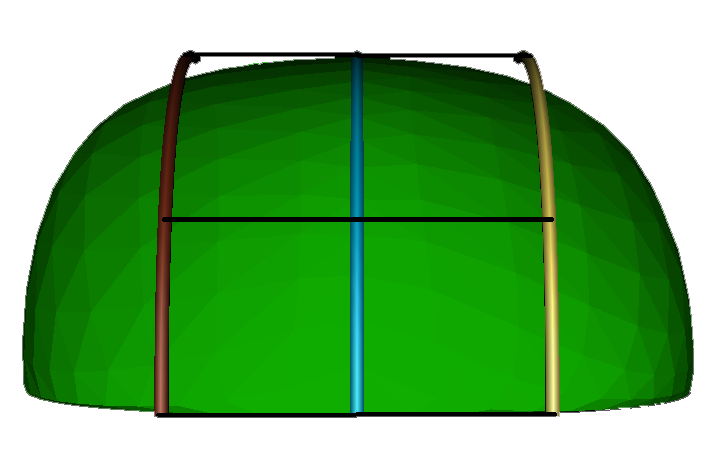}
\caption{A tangential ribbon is formed as a lofted boundary curve towards the direction of the tangents of the the two adjoining curves. Black line indicate that the ribbons and the boundary curve are co-linear. Shaded version to show the underlying triangulation.}
\label{fig:images/ribbon1.png}
\end{figure}

\begin{figure}[!ht]\centering
\includegraphics[width=\linewidth]{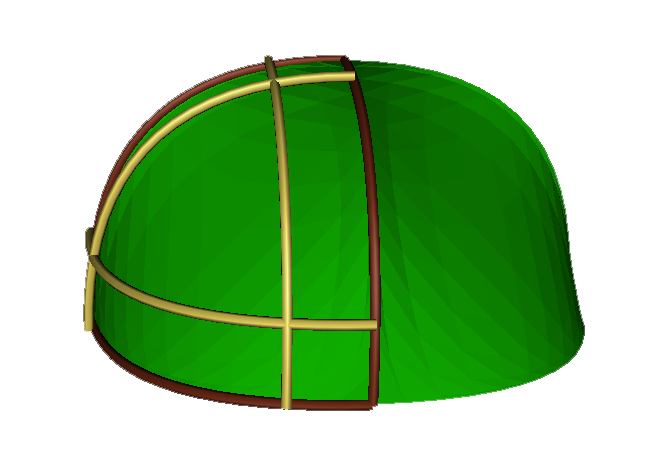}
\caption{A complete set of ribbons for two neighboring multi-sided patches, ribbons in the plane xy define tangent planes across the common boundaries and that yields G$^1$ continuity.}
\label{fig:images/ribbon2.png}
\end{figure}

Every boundary curve has a ribbon. Ribbons are constructed by taking the two tangents of adjoining curves and using them to offset the curve. The surface will approach the curve with the same slope as the ribbon if the two ribbons on a curve (one for each patch) are cotangent, then the surface will be. Furthermore, if the ribbons are the same interpolant of the tangents then one just has to make the tangents collinear to get $G^1$ across boundaries. An illustration of a ribbon of two neighboring three-sided surfaces is in figure \ref{fig:images/ribbon1.png}. Figure \ref{fig:images/ribbon2.png} shows a complete set of ribbons for a surface between two contiguous three-sided surfaces.

Adding ribbon information to the previous multi-surface formulation (\ref{eq:fundDef}), requires a distance function $s_i = d_i(u)$ from the set of points $u_i$ to the footprint $f_i$.$g_i$ are the lofted cubic Hermite splines defining the ribbons. Finally, the weight function is squared to satisfy continuity requirements, as stated in \cite{gao2005multi}. The new formulation is as follows (\ref{eq:fundDef2}):

\begin{align}
F(u)=\sum_{i}R_i(s_i,t_i)\hat{W}_i(u)^2 \label{eq:fundDef2}\\
{R_i}(s_i,t_i)=(1-s_i)f_i(t_i)+s_ig_i(t_i)
\end{align}

\begin{figure}[!ht]\centering
\includegraphics[width=\linewidth]{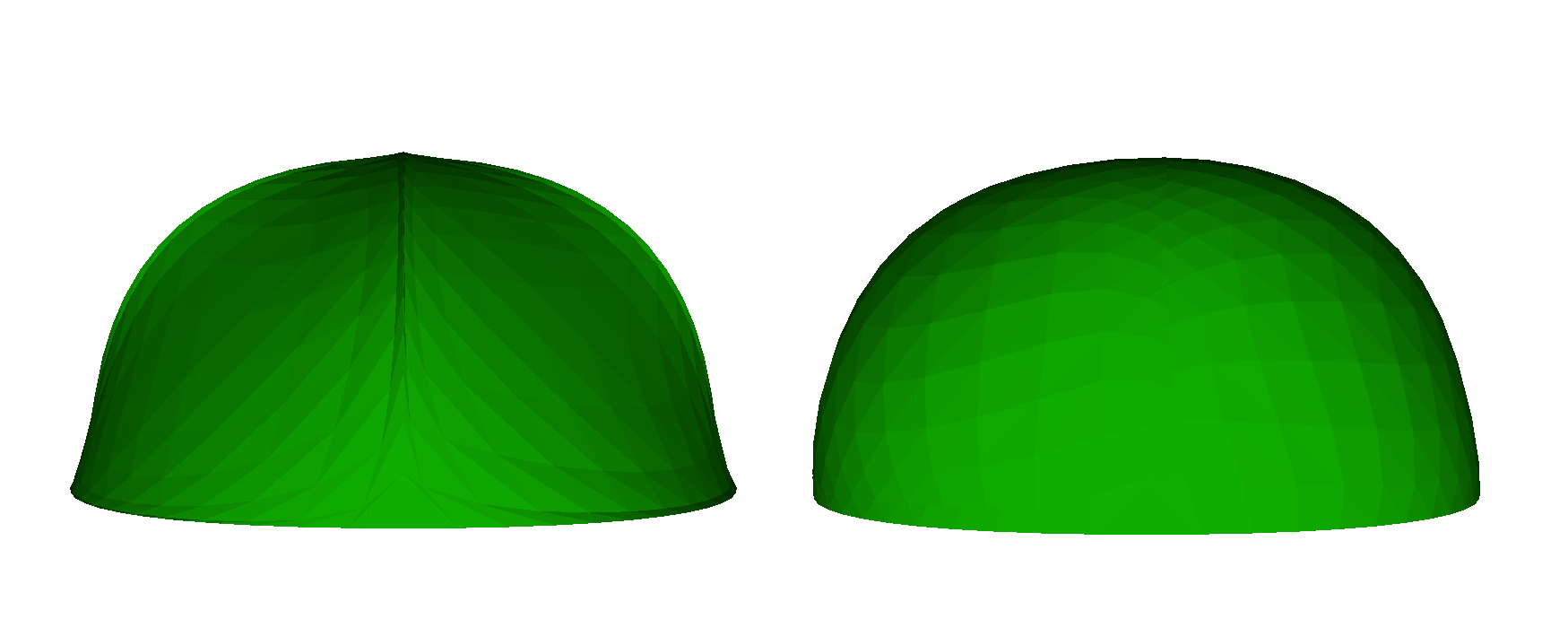}
\caption{Two neighboring three-sided patches, with G$^0$ and G$^1$ continuity respectively. Low resolution surfaces disclose the underlying triangularization.}
\label{fig:images/ribbonForm.png}
\end{figure}

The following figure shows two adjacent tree-sided surfaces with the previous G$^0$ formulation and the just stated G$^1$ formulation using tangential ribbons; a considerable upgrade on surface approximation can be noticed, see figure (\ref{fig:images/ribbonForm.png}).

\section{Results}

A set of implicit functions are here exposed to demonstrate the output of the marching surfaces algorithm. Isosurfaces constructed out of three, four, five and six-sided surfaces make up the final composite surface of the contour of interest. Comparisons between the marching cubes and the G$^0$ and G$^1$ representations are shown.

\begin{figure*}[htp]
	\centering

	\includegraphics[width=.31\textwidth]{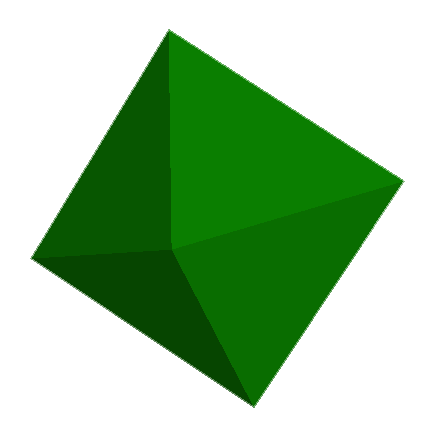}\hfill
	\includegraphics[width=.31\textwidth]{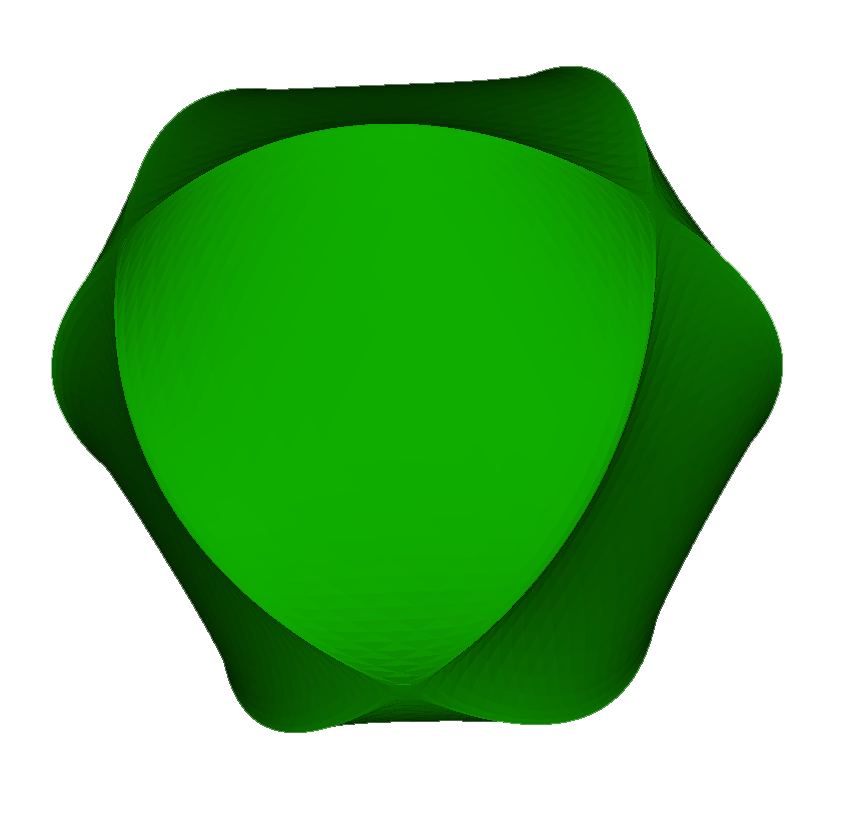}\hfill
	\includegraphics[width=.31\textwidth,keepaspectratio]{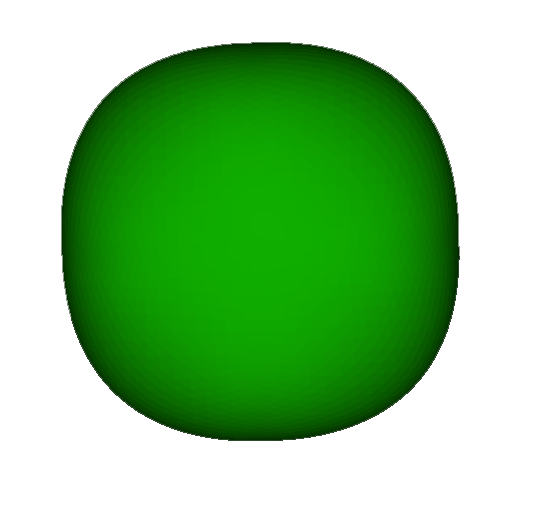}

	\caption{Scalar field $s(x,y,z) = x^2+y^2+z^2 -1 = 0 $ over [-3, -3, -3] x [3, 3 ,3], isovalue = 0. From left to right: (a) Marching cubes. (b) Multi-sided G$^0$ contours. (c) Multi-sided G$^1$ contours.}
	\label{fig:test1caseA}
\end{figure*}

Figures \ref{fig:test1caseA} shows the scalar field  $ s(x,y,z)=x^2+y^2+z^2 -1 = 0 $ over a [-3, -3, -3] x [3, 3 ,3] grid and isovalue = 0. Figure \ref{fig:test1caseA}a shows the isosurface extraction by the traditional marching cubes algorithm composed of 8 triangles. Figure \ref{fig:test1caseA}b shows the same scalar field approximation using marching surfaces with 8 three-sided surfaces and G$^0$ continuity. Figure \ref{fig:test1caseA}c shows the same isosurface using ribbons with G$^1$ continuity.

\textbf{Conclusions}. We have presented a method for approximating and rendering isosurfaces for implicit functions or scientific datasets. Our algorithm generates composite surfaces out of three, four, five and six possible sides. Visual appealing is directly related with the degree of smoothness of the approximation, and we described two alternatives that yield to $G^0$ and $G^1$ continuity across neighboring edges; both alternatives can be computed independently, i.e. cell-by-cell. The ability to store large amounts of geometry based on multi-sided surfaces and Hermite splines yields to a compact and straight forward storage method that is amicable with restricted environments such as web browsers and mobile devices. Large datasets are also good candidates for applications of the method here presented combining parallel paradigms and out-of-core techniques. These are all possible avenues for future work.

\newcommand{\etalchar}[1]{$^{#1}$}

\end{document}